\documentclass[]{article} 

\makeatletter
 
\PassOptionsToPackage{hyperref,svgnames,table}{xcolor}
\PassOptionsToPackage{T1}{fontenc}

\usepackage{algorithm}
\usepackage{amsmath}
\usepackage{amssymb}
\usepackage{booktabs}

\usepackage{fontenc}
\usepackage{graphicx}
\usepackage{hyperref}
\usepackage{ntheorem} 
\newtheorem{definition}{Definition} 
\usepackage{paralist}
\usepackage{ragged2e}
\usepackage[style=default]{subcaption}
\usepackage{tabularx}
\usepackage{wrapfig}
\usepackage{xcolor}

\usepackage{cleveref}

\newcommand{\HY}{\hyphenpenalty=25\exhyphenpenalty=25}
\newcolumntype{R}[1]{>{\hsize=#1\hsize\HY\RaggedLeft\arraybackslash\hspace{0pt}}X}
\definecolor{gsngreen}{rgb}{.84,.91,.83}
\AtBeginDocument{%
  \def\doi#1{\url{https://doi.org/#1}}}

\newcommand\mbPEhHP[1][]{\mathcal{D}_{#1}} 
\newcommand\mQghoDv[1][]{\mathcal{U}_{#1}} 
\newcommand\mQOELwE[1][X]{\mathcal{#1}}
\newcommand\mLTJYoy[1][]{#1^{\star}} 
\newcommand\muyxJqs[1]{|#1|}
\newcommand\mmWQIIu[1]{#1^\intercal} 
\newcommand\mBGFxda[1]{#1^+}
\def\mxXzOyg#1{\lVert#1\rVert}
\newcommand\mmoSmMG[1][X]{\mathcal{C}(#1)}
\newcommand\moCeHwO[1]{2^{#1}} 
\newcommand\mnWueTT[1][X]{\mathcal{T}(#1)}
\newcommand\mkfGRXG[1]{[\![#1]\!]}
\newcommand\mIHrBuo[1][]{V^{#1}}
\newcommand\mHtiAEm[1]{\mathbf{#1}}
\newcommand\mmbQYVt[1][]{W^{#1}}

\newcommand\mmZLevr[1][f]{\operatorname{coll}_{\mathsf{#1}}}
\newcommand\mcAFThG[1][]{J^{#1}}
\newcommand\mNFBizE[1][\delta]{\Delta_{#1}}
\newcommand\mimLwVu[1][f]{\mathit{#1}}
\newcommand\mGJEKgb[1]{\mathsf{#1}}
\newcommand\mavlbEx[1]{\mathsf{#1}}
\newcommand\mlYEpIk[1][]{\mathcal{O}_{\mathsf{#1}}}
\newcommand\mOnFCuJ[1][x]{\bar{#1}}

\makeatletter
\begin{document}
\title{A Parametric 
  Model for Near-Optimal Online Synthesis with Robust Reach-Avoid
  Guarantees
}
\author{Mario Gleirscher 
  \and Philip H\"onnecke} 
\maketitle
\label{l:1}

\begin{abstract}
  \emph{Objective:} To obtain explainable guarantees 
  in the online synthesis of optimal controllers for high-integrity
  cyber-physical systems, 
  we re-investigate the use of exhaustive search as an alternative to
  reinforcement learning.
  \emph{Approach:} We model an application scenario as a hybrid game automaton,
  enabling the synthesis of robustly correct and near-optimal
  controllers online without prior training.
  For modal synthesis, we employ discretised 
  games 
  solved via scope-adaptive and step-pre-shielded discrete dynamic programming. 
  \emph{Evaluation:} In a simulation-based experiment, we apply our
  approach to an autonomous aerial vehicle scenario.
  \emph{Contribution:} We propose a parametric system model and a
  parametric online synthesis.
\end{abstract}

\section{Introduction}
\label{l:2}

\paragraph{Motivation.}

\begin{figure}[t]
  \includegraphics[width=.65\linewidth,trim=250 35 40 0,clip]{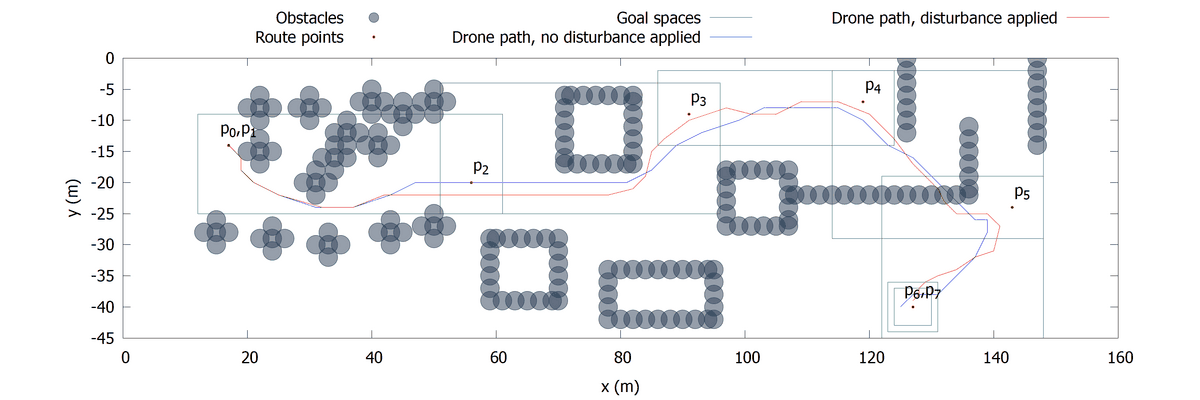}%
  \hfill
  \includegraphics[width=.33\linewidth,trim=250 0 150 100,clip]{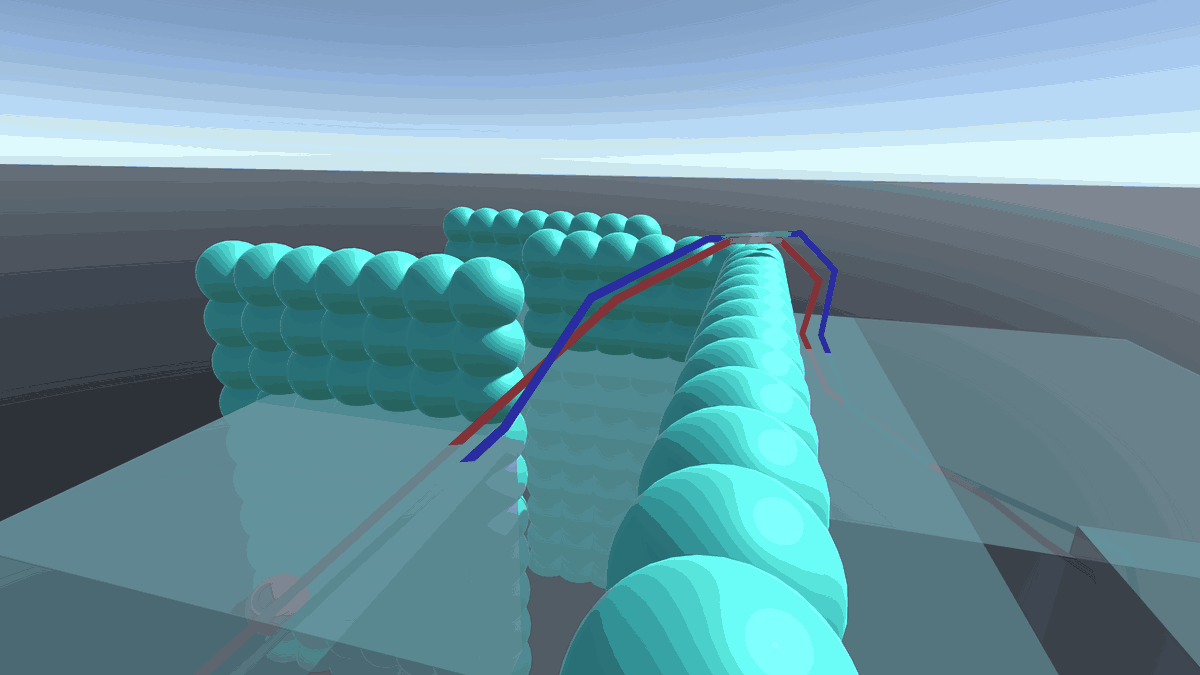} 
  \caption{Scene model for an AAV following a route ${\bar{r}}$
    from waypoint ${\mHtiAEm{p}}_0$ to ${\mHtiAEm{p}}_7$}
  \label{l:4}
\end{figure}

To achieve complex and critical tasks, systems 
with a high grade of autonomy perform decision making and control at
strategic and tactical levels under sparse human-machine interaction.  Consider, for
example, navigation of autonomous aerial vehicles~(AAVs), as illustrated in
\Cref{l:4}.
At the tactical level~(e.g., route segment tracking), often control
problems of non-linear, disturbed dynamical systems have to be solved
by constructing provably robust~(i.e., for safety) and
near-optimal~(i.e., for minimum-cost reachability)
controllers.  To accommodate uncertainties (e.g., changing
environments) and perform at scale, this type of controller synthesis
is preferably done online, that is, during operation and 
right before the actual use of the controllers.

\paragraph{Challenges.}

Reinforcement learning~(RL), a common and versatile type of approximate dynamic programming~(ADP) frequently used in
online synthesis, has shown to have several assurance-related
drawbacks:
\begin{itemize}
\item partial, imprecise, or even missing guarantees (due to a lack of
  behavioural coverage or an inappropriate initialisation
  \cite[Sec.~6]{b:4}),
\item post-hoc explanations (i.e., identifying failure causes may
  require training even with transparent algorithms
  \cite{b:10}), 
\item lack of robustness to changing operational profiles~(e.g.,
  obstacle scenarios 
  not covered or differing from training setups
  \cite[p.~30]{b:18}), and
  
\item limited or costly training (off-line/simulation or on-line/real
  interaction; requires active supervision, e.g., real-time
  safety 
  shielding \cite{b:20,b:13}).
\end{itemize}  
These limitations together with a need for more precise and complete
guarantees revive exhaustive, hence, less efficient forms of
synthesis.  However, to keep computation within bounds while
guaranteeing critical 
properties, online synthesis, whether or not based on RL, implies
restrictive trade-offs~(e.g., predictive imprecision, latencies in
segment tracking).

\paragraph{Research Question.} 

Not focusing on real-time performance, what kind of robust reach-avoid
guarantees can exhaustive numerical algorithms for near-optimal online
synthesis provide and under which assumptions?

\paragraph{Approach and Application.}
\label{l:3}

\begin{figure}
  \centering
  \footnotesize
\includegraphics[width=\linewidth]{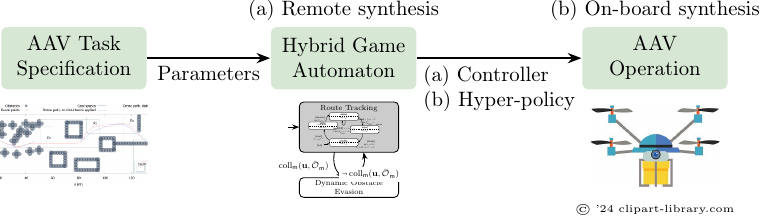}%
  \caption[Assured online synthesis]{Assured online synthesis with 
    (a) remote or (b) on-board computation}
  \label{l:5}
\end{figure}

We focus on an autonomous aerial vehicle~(AAV)-based delivery scenario
(\Cref{l:5}) formalised as a hybrid game automaton~(HGA).  The HGA
enables tactical controller synthesis to be done online (i.e., during
operation) via discrete dynamic programming~(DDP).  We permit a perforated\footnote{Start and
  end of a given route must be connected with a wide enough tube.
} fixed obstacle cloud and bounded uncertain wind disturbance.  Unsafe
actions possibly enabled by the discretisation are filtered during
DDP before finishing a search stage, which corresponds to
pre-shielding.  A
\emph{supervisor}~\cite{b:5},
a high-level controller, coordinates the interruption of the tactical
controller by an obstacle evasion unit if moving obstacles are about
to intrude the planned trajectory.

\paragraph{Related Work.}
\label{l:6}

Online synthesis of controllers for autonomous systems has been a
research subject for many decades, with a steady activity around
robustly correct, near-optimal control.  Below, we highlight some more
recent works.

\emph{Real-time-capable online synthesis} techniques have been
proposed for \emph{special-purpose} control of single- and multi-agent
systems.  
  For example, Li et al.~\cite{b:15} apply a
fast model-predictive control scheme focusing on stabilisation around
a given continuous reference trajectory.  They use a quadratic cost
term enabling initialised, sequential quadratic programming.
For AAV collision avoidance, Bertram et
al.~\cite{b:2} and Taye et
al.~\cite{b:19} realise multi-agent
reachability-based 
state space reduction to accelerate backward Markov decision process~(MDP) policy search.
In comparison, our approach is significantly slower but more widely
applicable, allowing non-convex cost terms, sparse reference
trajectories, near-optimality and guaranteed reach-avoidance under
bounded disturbance, and it is integrated into a flexible multi-tier
hybrid control scheme.

\emph{Scalable offline synthesis} techniques have been developed for
discrete and hybrid single- and multi-agent systems.
For example, Ivanov et
al.~\cite{b:12} apply
deep-RL to compute contract-compliant controllers for each mode
of a hybrid automaton. 
Gu et al.~\cite{b:8} use RL to
perform synthesis at scale for timed stochastic multi-agent systems. 
In \cite{b:17}, we refine a
stochastic Petri-net abstraction with partial state observability to
synthesise optimal schedules for tasked robot collectives.
By approximating fixpoints and sacrificing stochasticity in the
weighted reach-avoid game setting, we bypass an even higher complexity
of exact MDP synthesis.  This enables us to pre-process up to
200\,Mio.\ states in non-real-time, going far beyond the state space
in \cite{b:17} and yet in a
time scale similar to approaches, such as
\cite{b:8,%
  b:12}.  Our work
provides reach-avoid guarantees without a potentially intense training
needed for RL.  However, because of the scenario-sensitive
parameterisation, our technique can currently not be used in fast
real-time settings, such as autonomous driving or for collision
avoidance~\cite{b:15,%
  b:12}.

\paragraph{Contributions.}

Our approach enhances previous work as follows: 
\begin{enumerate}[(i)]
\item \emph{Step-shielded online synthesis:} We provide an 
  algorithm for solving discretised modal games with reach-avoid
  winning conditions.  The algorithm uses step-shielded DDP and
  supports non- and quasi-stationary policies (increasing robustness
  to disturbances 
  at the cost of performance) 
  \cite{b:14}.  
  We slightly reduce DDP's curse-of-dimensionality problem by
  scope adaptation\footnote{The relevant fraction of the state space
    is selected according to the current system state and extended
    on-the-fly to increase solvability of the online synthesis
    problem.} and fixpoint approximation for the winning region.
\item \emph{Parametric hybrid game model:} We construct a parametric
  weighted HGA covering a range of typical AAV scenarios.
  The automaton enables a three-tier separation to allow a trading off
  of operational cost and flexibility: higher-level supervision (e.g.,
  moving obstacle evasion), near-optimal trajectory tracking, and
  sub-tactical 
  control 
  enabling local policy learning (e.g., via deep-RL). 
  We embed our algorithm into a player for the HGA.
\end{enumerate}
A preliminary variant of (i) and (ii) was implemented and evaluated
in~\cite{b:11}.

\paragraph{Outline.}

After the preliminaries (\Cref{l:7}), we describe
our AAV application and control problem
(\Cref{l:10}).  Our contributions to online synthesis
with guarantees follow in the \Cref{l:21}.  We
evaluate our approach experimentally (\Cref{l:27})
and close with a discussion and remarks~(\Cref{l:33,%
  l:36}).

\section{Preliminaries}
\label{l:7}

\paragraph{Notation.} 

For a set of variables
$\mathit{Var}=X\uplus U\uplus D$,
let ${\mathbb{X}}\subset\ensuremath{\mathbb{Z}}^{X}$ be an $X$-typed,
finite,
$m$-dimensional Euclidean \emph{state space},\footnote{For a 3D
  coordinate $\mHtiAEm p$, we use the typical naming convention
  $\mHtiAEm p=\mmWQIIu{(x,y,z)}$.}  and let $\mQghoDv$ and $\mbPEhHP$
be $U$- and $D$-typed \emph{control} and
\emph{disturbance} ranges, each including $\mHtiAEm 0$.
Moreover, let 
the classes of terms $\mnWueTT[\mathit{Var}]$ 
and constraints
$\mmoSmMG[\mathit{Var}]$ over $\mathit{Var}$.
$\mxXzOyg\cdot$ is the corresponding 2-norm, 
$A\oplus B=\{a+b\mid a\in A\land b\in B\}$ the Minkowski sum of
$A,B\subseteq{\mathbb{X}}$, and $A|_{\mathit{Var}'}$ the projection of
tuples 
in $A$ to variables in $\mathit{Var}'\subseteq\mathit{Var}$,\footnote{We
  omit set parentheses when referring to singleton sets 
  in subscripts.}  and $\dashrightarrow$ indicates a partial map.
Point-wise operators 
are lifted as usual.\footnote{For example, $\min,\max$ on vectors of
  sets are evaluated element-wise and return a vector of scalars.
  Negation of a set returns the set of negated elements.}  Moreover,
$\underline I$ and $\bar I$ denote the lower and
upper bounds of an interval $I\subseteq\ensuremath{\mathbb{R}}$ and
$\underline I..\bar I$ signifies that
$I\subseteq\ensuremath{\mathbb{Z}}$.
$\mkfGRXG\varphi_{\mathfrak{M}}$ denotes $\varphi$'s models in
domain~${\mathfrak{M}}$,\footnote{For example, the
  class of states, state pairs, sequences, or transition systems.}
formally,
$\mkfGRXG\varphi_{\mathfrak{M}} = \{M\in{\mathfrak{M}}\mid M\models\varphi\}$.
Given the finite sequences $\mLTJYoy[{\mathbb{X}}]$ over
${\mathbb{X}}$, the length $\muyxJqs\mOnFCuJ$ of a sequence
$\mOnFCuJ\in\mLTJYoy[{\mathbb{X}}]$, and its value
$\mOnFCuJ_k\in{\mathbb{X}}$ at position $k\in1..\muyxJqs\mOnFCuJ$,
we use
$\bar{\mathbb{X}}(\mNFBizE[]),\tilde{\mathbb{X}}\subset\mLTJYoy[{\mathbb{X}}]$ to
denote the classes of \emph{trajectories} with any two subsequent states
inside some $\mNFBizE[]\subseteq\ensuremath{\mathbb{Z}}^m$ and with
$\tilde{\mathbb{X}}=\bar{\mathbb{X}}([\pm 1]^m)$.

\paragraph{Weighted Hybrid Game Automata.}

Given a set of modes $Q$,
an 
alphabet~$A$, events
$E\subseteq
Q\times A\times Q$, and the hybrid
state space ${\mathcal{S}} = Q\times{\mathbb{X}}$,
a weighted HGA
$G =
(\mathit{Gra},\mathit{Var},\mathit{Ini},\mathit{Inv},\mimLwVu,\mathit{Jmp},F)$
comprises a
\begin{inparaitem}[]
\item mode transition graph
  $\mathit{Gra} = (Q, A,$ $E)$,
\item initial conditions $\mathit{Ini}\colon Q\to\mmoSmMG$,
\item invariants
  $\mathit{Inv}\colon Q\to\mathcal{C}(X)$,
\item flow constraints
  $\mimLwVu\colon
  {\mathcal{S}}\to\mmoSmMG[\mathit{Var}\cup\dot{X}]$,
\item jump conditions
  $\mathit{Jmp}\colon E\to\mmoSmMG[\mathit{Var}\cup
  \mBGFxda{X}]$
  comprised of guards and updates, and
\item weighted 
  winning conditions
  $F\colon{\mathcal{S}}\to\moCeHwO{\mnWueTT[\mathit{Var}]}$,
  generalising
  final conditions $\mathit{Fin}\colon Q\to\mmoSmMG$
  \cite{b:9}.
\end{inparaitem}
The copies $\dot{X}$ and $\mBGFxda{X}$ refer to the time
derivatives and discrete updates of $X$.  $\operatorname{grd}(e)$ and
$\operatorname{upd}(e)$ denote the guard and update conditions of
$\mathit{Jmp}(e)$.  In the remainder, we frequently use
${\mHtiAEm{s}}=(q,{\mHtiAEm{x}})\in{\mathcal{S}}$ for a state
of~$G$.

\paragraph{Integer Difference Games.}

Let
$\Pi_{\mQOELwE[Y]}={\mathbb{X}} 
\times\mathbb{N}_+\dashrightarrow\mQOELwE[Y]$
be a \emph{strategy space} for a player with alphabet~$\mQOELwE[Y]$.
Given a flow constraint~$\mimLwVu$
and its discretised Taylor
expansion~$\hat\mimLwVu$,
we associate a state ${\mHtiAEm{s}}$ of~$G$ with a discretised
two-player game
$G_{\mHtiAEm{s}}=(\mathcal{X},
\hat\mimLwVu,F)$.\footnote{Players share the state and time but
  do not know each others' future actions.}  We use
${\mHtiAEm{u}}\colon\Pi_{\mQghoDv}$ and
${\mHtiAEm{d}}\colon\Pi_{\mbPEhHP}$ to describe
the 
memory-less 
strategy profile of the two players, the controller and the
environment, based on the
non-empty 
control and disturbance ranges $\mQghoDv$ and $\mbPEhHP$.  Then,
for a horizon $N$ and the winning condition
$F=(L,\Phi)$ with
stage 
and terminal costs~$L$ and~$\Phi$, a
finite-horizon, 
discrete optimal 
controller~${\mHtiAEm{u}}^*\colon\Pi_{\mQghoDv}$ can be obtained
from solving a constrained, discrete dynamic optimisation
problem~\cite{b:3}.  Such a
problem can be solved with a forward-Euler DDP (\Cref{l:9})
for the period $I=1..N$.

\begin{algorithm}
  \caption{Discrete dynamic programming} 
  \label{l:9}
\includegraphics[width=\linewidth]{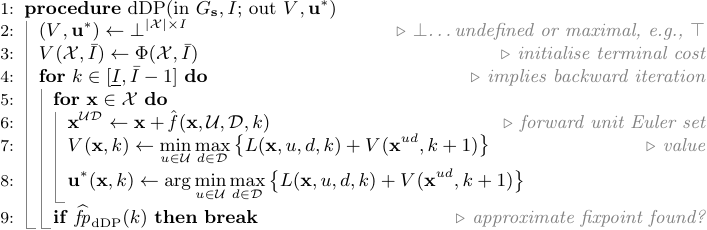}%
\end{algorithm}
 
The single-step successor
${\mHtiAEm{x}}^{ud} 
= {\mHtiAEm{x}} +
\hat\mimLwVu({\mHtiAEm{x}},u,d,k)%
$ 
is lifted in
Line~6, such that
${\mHtiAEm{x}}^{ud}\in
{\mHtiAEm{x}}^{\mQghoDv\mbPEhHP}\subseteq{\mathbb{X}}$.
Moreover, let
${\mHtiAEm{x}}^{{\mHtiAEm{u}}{\mHtiAEm{d}}}\in\bar{\mathbb{X}}$ be the
$N$-step successor (trajectory) 
of $G$ emanating from ${\mHtiAEm{x}}\in{\mathbb{X}}$ under
influence of $({\mHtiAEm{u}},{\mHtiAEm{d}})$, and
${\mHtiAEm{x}}^{{\mHtiAEm{u}}\mbPEhHP}\subseteq\bar{\mathbb{X}}$ be the family of
such trajectories 
under~$\mbPEhHP$.
$\mIHrBuo({\mHtiAEm{x}},k)\in[0,\top]$
with $\top<\infty$ is the finite-horizon
value corresponding to the optimal cost-to-go, using the optimal
profile~$({\mHtiAEm{u}}^*,{\mHtiAEm{d}}^*)$.
Line~9 allows a check
$\widehat{\mathit{fp}}_{\mathrm{dDP}}$ for whether a fixpoint is approximated
prior to reaching~$N$.  While the actual fixpoint yields a
stationary 
optimal strategy profile, 
a fixpoint approximation still guarantees bounded correct
(quasi-stationary and near-optimal) strategies under reduced memory
consumption and computational effort.  Let $\mathcal{G}$ be the class of
integer difference games.

\section{Aerial Delivery as a Hybrid Game}
\label{l:10}

\begin{figure}
  \centering
  \small
\includegraphics[width=\linewidth]{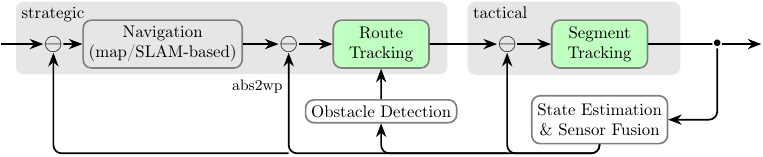}%

   \caption{Overall AAV control block structure}
  \label{l:11}
\end{figure}

\paragraph{Assumptions.}

\begin{figure}
  \subfloat[Parametric tactical control logic {$G[\mHtiAEm{U},T]$}]{
    \label{l:13}
    \footnotesize
\includegraphics[width=.55\linewidth]{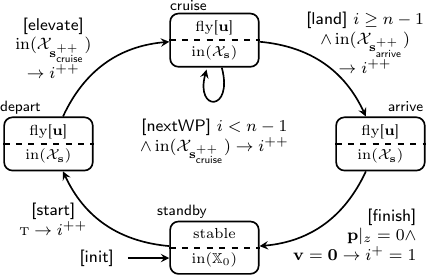}%

  }
  \subfloat[Supervisor logic]{
    \label{l:14}
    \footnotesize
\includegraphics[width=.35\linewidth]{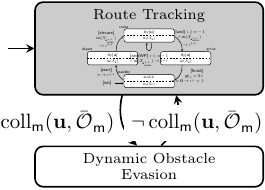}%

 }%
  \caption{Tactical AAV control ($\mimLwVu$/$\mathit{Inv}$ are
    above/below the dashed lines in the modes, and
    $\mathit{Jmp}$ is shown in the transition labels) and supervisor
    logic}
  \label{l:15}
\end{figure}

Let $X=\{{\mHtiAEm{p}},{\mHtiAEm{v}},i\}$ with the AAV
\emph{position}\footnote{${\mHtiAEm{p}}|_z$ measures distance above local
  ground outside buildings.}  ${\mHtiAEm{p}}\colon\ensuremath{\mathbb{Z}}^3$ and
\emph{velocity} ${\mHtiAEm{v}}\colon[\pm v_{\max}]^3$ (with an absolute
maximum at $v_{\max}$) and a \emph{next-waypoint} index
$i\colon\mathbb{N}$.  We use convex 
sets $\mathcal{X}_{\mHtiAEm{s}}\subseteq{\mathbb{X}}$, called
\emph{scopes},\footnote{$\mathcal{X}_{\mHtiAEm{s}}$ is not included as a
  state variable in $X$ because it can be derived from
  ${\mHtiAEm{x}}$.}  and $\mathcal{P}={\mathbb{X}}|_{\mHtiAEm{p}}$, the
position 
grid of the \emph{scenery}.
Given ${\mathbb{X}}$, an AAV \emph{task}
$T = ({\mathbb{X}},{\bar{r}},\mlYEpIk)$ comprises a \emph{fixed 
  obstacle cloud} $\mlYEpIk\subset\mathcal{P}$ and a \emph{route}
${\bar{r}}=\{{\mHtiAEm{p}}_i\}_{i=1}^n$ where the predicate
$\operatorname{valid}({\bar{r}})$ requires ${\bar{r}}$ to have $n\geq 4$
different 
\emph{waypoints} ${\mHtiAEm{p}}_i\in\mathcal{P}$ with ${\mHtiAEm{p}}_{1,n}$ being on
the ground, ${\mHtiAEm{p}}_{2,n-1}$ $(x,y)$-superimposed, and ${\mHtiAEm{p}}_{2..n-1}$
residing above a minimum height~$z_{\min}$.
Given a cube 
$\mNFBizE=[\pm\delta]^3$ for a small
$\delta\in\mathbb{N}_0$, 
we require~$\mlYEpIk$ to be $\delta$-\emph{perforated}, such that
\begin{align}
  \label{l:12}
  \exists \mOnFCuJ\in\tilde{\mathbb{X}}\colon
  {\bar{r}}\subseteq
  \underbrace{\mOnFCuJ|_{{\mHtiAEm{p}}}\oplus\mNFBizE}_{\text{cont.\ $\delta$-tube}}
  \land
  \mOnFCuJ|_{{\mHtiAEm{p}}}\oplus\mNFBizE 
  \cap\mlYEpIk=\varnothing\,.
\end{align}
Informally, the route is enclosed in a 
tube with radius $\delta$ not colliding with fixed
obstacles. 
We assume $({\bar{r}},\mlYEpIk)$ to be delivered by map- and SLAM-based
navigation and sensor fusion on-board the
AAV~(\Cref{l:11}) and allow
$({\bar{r}},\mlYEpIk)$ to be updated at every~${\mHtiAEm{p}}_i$ invariant
under \eqref{l:12} and $\operatorname{valid}({\bar{r}})$.\footnote{In
  \Cref{l:15} and below, we use predicates over
  $\mlYEpIk,\mlYEpIk[m],{\bar{r}}$, and ${\mHtiAEm{u}}$.  $\mlYEpIk$ and
  ${\bar{r}}$ are game parameters, ${\mHtiAEm{u}}$ is a parameter of the
  tactical control's alphabet (\Cref{l:21}),
  and $\mlYEpIk[m]$ and ${\mHtiAEm{u}}$ are parameters of the supervisor's
  alphabet.}

\paragraph{Tactical Control.}

We consider a 
HGA $G$ (\Cref{l:7}) combining the logic
of the AAV route and segment tracking units, where $\mathit{Gra}$,
$\mathit{Inv}$, $\mimLwVu$, and $\mathit{Jmp}$ are illustrated in
\Cref{l:13} and
$\mathit{Ini}(q)=\textsc{t}$ for $q\in Q$.
For brevity, \eqref{l:12} and $\operatorname{valid}({\bar{r}})$ as
global invariants remain implicit in $\mathit{Ini}$ and $\mathit{Inv}$.  To
compute the current scope, we use the function
\begin{align}
  \label{l:16}
  \mathcal{X}_{\mHtiAEm{s}} =
  \begin{cases}
    ({\mHtiAEm{p}}_i,\mHtiAEm 0) \oplus
    \big(
    [\pm\delta_{p,q}]^2\times[0,\max\{{\mHtiAEm{p}},{\mHtiAEm{p}}_i\}
    \big.
    &q\in\{\mavlbEx{depart},
    \\\quad
    +\delta_{p,q}]
    \times[\pm\delta_{v,q}]^3
    \big)
    \times\{i\},
    &\quad\mavlbEx{arrive},\mavlbEx{standby}\}
    \\
    \big([\min\{{\mHtiAEm{p}},{\mHtiAEm{p}}_i\},\max\{{\mHtiAEm{p}},{\mHtiAEm{p}}_i\}]
    \times
    \{\mHtiAEm 0\}\big)\oplus
    &q\in\{\mavlbEx{cruise}\}
    \\\quad
    \big([\pm\delta_{p,q}]^3
    \times[\pm\delta_{v,q}]^3
    \big)
    \times\{i\}
  \end{cases}
\end{align}
describing \emph{transition cuboids} (\Cref{l:4} left and
right) around route segments 
for moments when the AAV is near one waypoint and proceeds to the
next.
We use $\delta_{p,q}$ and $\delta_{v,q}$ for position
and velocity padding of mode $q$ and the abbreviations
$\operatorname{in}(\mathcal{X}) \equiv {\mHtiAEm{x}}\in\mathcal{X}$,
${\mathbb{X}}_0 \equiv \{{\mHtiAEm{x}}\in{\mathbb{X}}\mid
{\mHtiAEm{p}}|_z=0\land{\mHtiAEm{v}}=\mHtiAEm 0\land i=1\}$,
${\mHtiAEm{s}}^{++}_q = (q,\mmWQIIu{({\mHtiAEm{p}}_i,\mHtiAEm
  0,i+1)})$,\footnote{Note how ${\mHtiAEm{s}}^{++}_q$ refers
  to the next waypoint ${\mHtiAEm{p}}_i$ and the one after that,
  ${\mHtiAEm{p}}_{i+1}$.  ${\mHtiAEm{s}}^{++}_q$ will be used to
  compute the goal region for a modal game.}
and
$i^{++}\equiv\mBGFxda{i}=i+1$.

The AAV dynamics\footnote{At tactical control level, we employ a
  simplification of the dynamics to a point mass.}  in all modes,
except for $\mavlbEx{standby}$, is given by
\begin{align}
  \label{l:17}
  \mimLwVu[fly]
  = 
  \dot{{\mHtiAEm{x}}} =
  \begin{bmatrix}
    \dot{\mHtiAEm{p}}\\
    \dot{\mHtiAEm{v}}\\
    \dot{i}
  \end{bmatrix}
  =
  \begin{bmatrix}
    {\mHtiAEm{v}}\\
    {\mHtiAEm{u}}+{\mHtiAEm{d}}+\mHtiAEm g\\
    0
  \end{bmatrix}
\end{align}
with an implicit unit mass (i.e., vehicle + payload = 1),
an approximate\footnote{$\mHtiAEm g$ is omitted in the isolated dynamics.
  The compensation of numerical imprecision can be delegated to
  lower-level stability control based on system identification.}
gravitational force $\mHtiAEm g=\mmWQIIu{(0,0,-10)}$, and
${\mHtiAEm{u}}\in\mQghoDv$, ${\mHtiAEm{d}}\in\mbPEhHP$ for
$\mQghoDv,\mbPEhHP \colon \moCeHwO{\ensuremath{\mathbb{Z}}^3}$.  In
$\mavlbEx{standby}$ mode, the flow condition is
$\mimLwVu[stable]\equiv\dot{{\mHtiAEm{x}}}=\mHtiAEm 0$.

In our experiments, we use small
$\mQghoDv,\mbPEhHP\subseteq[\pm 2]^3$
to allow the AAV to
accelerate in 26 directions and wind disturbance to occur in 4
directions, as well as $v_{\max}=10$ ranging down to 5 for
efficiently handling smaller scopes.

\paragraph{Supervisory Control.}

Strategic AAV control comprises a
supervisor~(\Cref{l:14}), which separates route and
segment tracking from moving-obstacle evasion.  The supervisor
interrupts tactical control and resumes it after an evasion manoeuvre.
Encoded in $F$, for all modes but
$\mavlbEx{standby}$, the unsafe set encompasses (inevitable) collisions
with fixed obstacles as\footnote{With
  $\mxXzOyg{\mHtiAEm o-{\mHtiAEm{p}}}\leq\delta$ instead of
  ${\mHtiAEm{p}}\in\mHtiAEm o\oplus\mNFBizE$, we can work with
  $\delta=1.5$ in experiments.}
\begin{align}
  \label{l:18}
  \mmZLevr(\mlYEpIk) 
  \equiv
  \exists\mHtiAEm o\in\mlYEpIk\colon
  \mxXzOyg{\mHtiAEm o-{\mHtiAEm{p}}}\leq\delta%
  ,\quad\text{if}\;q\neq\mavlbEx{standby}.
\end{align}
For a set $\tilde{\mlYEpIk[]}_{\mathsf{m}}$ 
of trajectories of moving obstacles tracked by the supervisor,
\begin{align*} 
  \mmZLevr[m]({\mHtiAEm{u}},\tilde{\mlYEpIk[]}_{\mathsf{m}})
  &\equiv
    \exists k\in I, 
    \bar o\in\tilde{\mlYEpIk[]}_{\mathsf{m}}\colon
    {\mHtiAEm{x}}^{{\mHtiAEm{u}}\mbPEhHP}_k|_{{\mHtiAEm{p}}}
    \cap
    \bar o_k\oplus\mNFBizE
    \cap
    {\mHtiAEm{p}}\oplus\mNFBizE[sbd]({\mHtiAEm{v}})
    \neq\varnothing
\end{align*}
indicates that the predicted trajectory~$\bar o$ of some moving
obstacle crosses the AAV trajectory
${\mHtiAEm{x}}^{{\mHtiAEm{u}}\mbPEhHP}|_{{\mHtiAEm{p}}}$ within safe braking
distance $\mNFBizE[sbd]$.
Evasion manoeuvres (i.e., replacing ${\mHtiAEm{u}}$ by an evasive
${\mHtiAEm{u}}_e$) could be computed by \Cref{l:9}.  However, there are
other ways of computing ${\mHtiAEm{x}}^{{\mHtiAEm{u}}\mbPEhHP}|_{{\mHtiAEm{p}}}$
efficiently (e.g., \cite{b:1,%
  b:20})
and ${\mHtiAEm{u}}_e$ with
guarantees~(e.g., \cite{b:16}).  To
simplify our setting, we assume
$\tilde{\mlYEpIk[]}_{\mathsf{m}}=\varnothing$ and refer to the broader
treatment of supervision in, for
example,~\cite{b:5}.

Let us now formulate the synthesis problem focused on in this work.

\begin{definition}[Online Synthesis Problem]
  \label{l:20}
  Given a hybrid game automaton~$G$ and a task~$T$, continuously
  find tactical controllers ${\mHtiAEm{u}}$ steering the system 
  along route~${\bar{r}}$ while safely circumventing unsafe
  regions. 
  For correctness, we need ${\mHtiAEm{u}}$ to be
  \begin{inparaenum}[(i)]
  \item $\delta$-robust~(i.e., safe under bounded disturbance
    $\mbPEhHP$),
  \item near-optimal~(i.e., follow the minimum-cost path inside
    padded segment scopes, approximating all waypoints), and
  \item reaching the endpoint of ${\bar{r}}$.
  \end{inparaenum}
\end{definition}

\section{Online Synthesis via Parametric Games}
\label{l:21}

\paragraph{Parametric Modal Games.}

We assume that $\mathit{Gra}$ is controllable,\footnote{All jumps are solely
  controlled or chosen by the system.}  such that $G$ reduces to a
parametric
reach-avoid 
integer 
difference game
$G_{\mHtiAEm{s}} = (\mathcal{X}_{\mHtiAEm{s}},
\hat\mimLwVu,F)$ 
to be solved for and played after a
jump to~${\mHtiAEm{s}}$.
The forward-Euler scheme in Line~6
of \Cref{l:9} uses an isolated\footnote{E.g.,
  \eqref{l:17} without $\mHtiAEm g$ and under a change of
  variables via
  ${\mHtiAEm{x}}_{\mathrm{iso}}={\mHtiAEm{x}}_{\mathrm{orig}} -
  \mmWQIIu{({\mHtiAEm{p}}_i,\mHtiAEm 0,0)}$. 
} variant of the right-hand side of
$\hat\mimLwVu$. 
For $F=(L,\Phi)$, we employ
\begin{align}
  \label{l:22}
  L({\mHtiAEm{u}},{\mHtiAEm{d}};{\mHtiAEm{s}},k) =
  \begin{cases}
    0,\;\text{if}\;\rho\land\neg\alpha
    \\
    \top,\;\text{if $\alpha$ (shielded)}
    \\
    \lambda({\mHtiAEm{u}},{\mHtiAEm{d}};{\mHtiAEm{x}},k)
    ,\;\text{otherwise}
  \end{cases}
  \Phi({\mHtiAEm{s}}) =
  \begin{cases}
    0,\;\text{if}\;\rho\land\neg\alpha\\
    \top,\; \text{otherwise}
  \end{cases}
\end{align}
where $\rho$ and $\alpha$ specify the goal and unsafe regions to
be reached and avoided, respectively.  We assume that
$\mkfGRXG\rho\subseteq\mathcal{X}_{\mHtiAEm{s}}$ and
$\mkfGRXG\alpha\subseteq{\mathbb{X}}$ with
$\mkfGRXG\rho\setminus\mkfGRXG\alpha\neq\varnothing$.
Furthermore,
$\lambda({\mHtiAEm{u}},{\mHtiAEm{d}};{\mHtiAEm{x}},k) = \mmWQIIu{\mHtiAEm{x}}
P{\mHtiAEm{x}} + \mmWQIIu{\mHtiAEm{u}} Q{\mHtiAEm{u}} + \mmWQIIu{\mHtiAEm{d}} R{\mHtiAEm{d}}$ is a
weight term with correspondingly dimensioned matrices
$P$, $Q$, and $R$.

$L$ penalises $\alpha$ and rewards $\rho$, maximally.
Given \eqref{l:12}, \eqref{l:22}
implies $\mIHrBuo(\mHtiAEm 0,\cdot)=0$.  Not shown here, $k$ can be
used in $\lambda$ to penalise run-time.
Overall, the call $\operatorname{dDP}(G_{\mHtiAEm{s}},I)$
(\Cref{l:9}) provides a near-optimal
controller~${\mHtiAEm{u}}^*_{\mHtiAEm{s}}$.\footnote{Deviating from
  \Cref{l:7}, we pass $q$ to dDP by calling
  $L$ and $\Phi$ with
  ${\mHtiAEm{s}}=(q,{\mHtiAEm{x}})$.}

\paragraph{Bounded Correctness via Approximate Value Fixpoints.}

Let 
$\mmbQYVt(\mathcal{X}_{\mHtiAEm{s}},k) =
\{{\mHtiAEm{x}}\in\mathcal{X}_{\mHtiAEm{s}}\mid
\mIHrBuo({\mHtiAEm{x}},k)<\top\}$ be
an 
approximation of $G_{\mHtiAEm{s}}$'s 
winning region at time $k$ when playing no more than
$N-k$ 
steps.  In particular,
$\mmbQYVt(\mathcal{X}_{\mHtiAEm{s}},N) =
\{{\mHtiAEm{x}}\in\mathcal{X}_{\mHtiAEm{s}}\mid \Phi({\mHtiAEm{x}})<\top\}$.
To reduce the computational effort in \Cref{l:9}, we employ a
condition $\widehat{\mathit{fp}}_{\mathrm{dDP}}$ for premature termination in
Line~9.  It is defined as
\begin{align}
  \label{l:23}
  \widehat{\mathit{fp}}_{\mathrm{dDP}}(k)
  \equiv
  \muyxJqs{\mmbQYVt(\mathcal{X}_{\mHtiAEm{s}},k)} 
  =
  \muyxJqs{\mmbQYVt(\mathcal{X}_{\mHtiAEm{s}},k+1)}
  \lor 
  \widehat{\mathit{fp}}_{\mHtiAEm{U}}(k)\,,
\end{align}
where the conjunct
\[
  \widehat{\mathit{fp}}_{\mHtiAEm{U}}(k)
  \equiv
  \exists k'\geq k\colon
  \top\not\in\mIHrBuo({\mHtiAEm{x}}\oplus\mNFBizE,k')
\]
ensures that, latest at step $k$, ${\mHtiAEm{x}}$ is
$\delta$-robustly located inside
$\mmbQYVt(\mathcal{X}_{\mHtiAEm{s}},k)$.  This approximation
allows us to check whether the number of states with a \emph{robustly
  safe and goal-reaching trajectory} stabilises at~$k$ and,
hence, within the horizon~$N$.
Fixpoint approximation safes time but it limits our approach to
bounded correctness and increases the deviation of the discrete
solutions from optimality.

\paragraph{Scope Adaptation.}
 
To further reduce computational effort, we use \Cref{l:9} with a
spatio-temporal \emph{scope extension}.
The result is
\Cref{l:24}, with the aim to increase solvability of
$G_{\mHtiAEm{s}}$ by checking
$\widehat{\mathit{fp}}_{\mHtiAEm{U}}(1)$ in
Line~7.

\begin{algorithm}
  \small
  \caption{DDP with scope extension (in
    Line~4, note that
    ${\mHtiAEm{s}}=(q,{\mHtiAEm{x}})$)}
  \label{l:24}
\includegraphics[width=\linewidth]{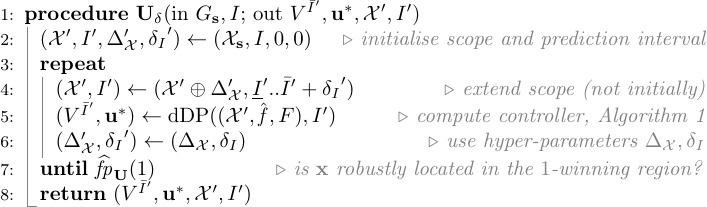}%
\end{algorithm}

\paragraph{Hyper-Policies and Pre-Computation.}
\label{l:25}

\Cref{l:24} realises a 
\emph{hyper-policy}
$\mHtiAEm{U}_{\delta}\colon\mathcal{G}\times\moCeHwO{\mathbb{N}_+}
\to({\mathbb{X}}\times\mathbb{N}_+\dashrightarrow\mQghoDv)$ for the parametric
$G[\mHtiAEm{U},T]$ in \Cref{l:13}.
${\mHtiAEm{u}}^*_{\mHtiAEm{s}} =
\mHtiAEm{U}_{\delta}(G_{\mHtiAEm{s}},I)$ is
defined for
$\mathcal{X}'\times I'\subset{\mathbb{X}}\times\mathbb{N}_+$ and
solves $G_{\mHtiAEm{s}}$ by determinising the
$(\mQghoDv,\mbPEhHP)$-underspecified modal 
dynamics of $G$.  ${\mHtiAEm{u}}^*_{\mHtiAEm{s}}$ is intended to be
computed online, remotely or on-board (\Cref{l:5}), if
${\mHtiAEm{x}}\in\mathit{Inv}(q)$.  Any suffix of ${\bar{r}}$ after
${\mHtiAEm{p}}_i$ can be updated 
together with the corresponding update of~${\mHtiAEm{u}}^*_{\mHtiAEm{s}}$. 

\paragraph{A 
  Hybrid Game Player.}

\Cref{l:26} combines \Cref{l:24}
with an execution routine for HGAs.  A play of
$G_{\mHtiAEm{s}}$ is a co-execution of both players successively
applying a strategy profile, presumably $({\mHtiAEm{u}}^*,{\mHtiAEm{d}})$, to
the dynamics~$\hat\mimLwVu$.
The loop
(Lines~3 to~10)
plays $G_{\mHtiAEm{s}}$ starting from
${{\mHtiAEm{x}}_0}={\mHtiAEm{s}}|_{\mHtiAEm{x}}$.  For the possible jump in
Line~5, the
$\exists$ of the \textbf{switch} is to be read as ``check and
pick~$e$''.\footnote{$G$ as a specification is agnostic to
  when and how guards are
  used. 
  We use guards as triggers because \Cref{l:26}
  interprets $G$ as a simulator.  Updates 
  will be performed as soon as a modal play enters the corresponding
  guard region.}  The numerical integrator in
Line~9 describes the simultaneous inputs
$({\mHtiAEm{u}}^*,{\mHtiAEm{d}})$ of both players being used in the dynamics and
adds the corresponding increment to the current state~${\mHtiAEm{x}}$,
resulting in a hybrid trajectory~$\bar{\mHtiAEm{s}}\in\bar{\mathcal{S}}$.

\begin{algorithm}
  \caption{%
    Hybrid game player (with restricted task updates)}
  \label{l:26}
\includegraphics[width=\linewidth]{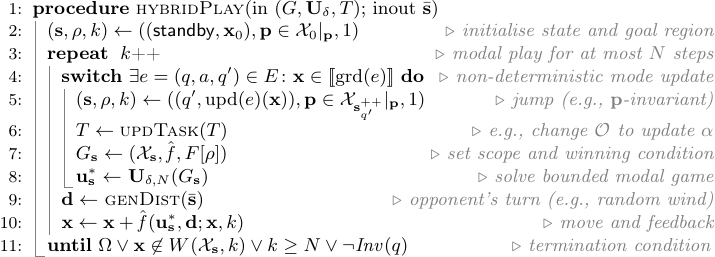}%
\end{algorithm}

Execution (Line~10) is
constrained by several conditions:
The first, $\Omega$, specifies \emph{termination}.
The other three,
${\mHtiAEm{x}}\not\in\mmbQYVt(\mathcal{X}_{\mHtiAEm{s}},k)$,
$k\geq N$, and $\neg\mathit{Inv}(q)$, specify
\emph{failure} (i.e., unsafe behaviour,
${\mHtiAEm{x}}\not\in\mkfGRXG{\rho\land\neg\alpha}$ if $k=N$),
\emph{timeout} (beyond $\bar I=N$, control choices
can no more rely on $\mIHrBuo({\mHtiAEm{x}},k)<\top$), and
\emph{invariant violation} (e.g., behaviour or an application of the
controller outside $\mathit{Inv}(q)$ is not specified).  These
three events, by definition, only occur if the current modal game
$G_{\mHtiAEm{s}}$ could not be solved.
In a simulator, these events can be used for model debugging, while
during operational tests, such events may serve, for example, system
identification and the adjustment of observers.

\paragraph{Liveness Considerations.}

Through \textsc{updTask}, \Cref{l:26} allows restricted
updates of $T$ during operation.  To avoid unrealistic
moving-target situations, we simplify our setting to a fixed global
state space ${\mathbb{X}}$ and route ${\bar{r}}$ for the entire
operation and to a fixed unsafe region $\alpha$ for each modal play.
This simplification provides a conservative assumption for achieving
strategic liveness.  The liveness proofs relying on these assumptions
are provided in a working
paper~\cite{b:6}.

\section{Implementation and Experiments}
\label{l:27}

We implemented the \Cref{l:9,l:24,l:26}
in {C\texttt{++} and evaluated their accuracy and performance in four AAV scenarios
(i.e., navigating
a small domestic \emph{yard},
a 200$\times$250\,$m^2$ \emph{industrial} area,
a 400$\times$450\,$m^2$ neighbourhood with several \emph{streets},
and a \emph{random} obstacle cloud) covering a range of realistic
situations.  With our simulator (\Cref{l:26}), we
illustrate plays of the hybrid game (\Cref{l:13})
for the \emph{yard} and \emph{industrial} scenarios
from~\cite{b:11} as well as the more
complicated \emph{streets} scenario.  We also highlight some data
(\Cref{l:32}).

\paragraph{Aerial Delivery Game Parameters.}

For the mentioned scenarios, we use
\begin{align}
  \label{l:28}
  \rho 
  &\equiv
    {\mHtiAEm{p}}\in\mathcal{X}_{{\mHtiAEm{s}}^{++}_{q'}}|_{\mHtiAEm{p}}
  &\alpha
  &\equiv\mmZLevr(\mlYEpIk)
  &\text{and}& 
  &\lambda({\mHtiAEm{u}},{\mHtiAEm{d}};{\mHtiAEm{s}},k)
  &={\mHtiAEm{x}}^2+{\mHtiAEm{u}}^2.
\end{align}
Informally, $\rho$ enforces the AAV to reach the next
waypoint segment $({\mHtiAEm{p}}_i,{\mHtiAEm{p}}_{i+1})$, except for the elevate
segment during departure, and $\alpha$ enforces the vehicle to avoid
static obstacles.  A corresponding collision check (i.e., safety
pre-shielding)
with $\mlYEpIk$ is performed for each state and pair of control and
disturbance inputs in Line~7 of
\Cref{l:9} and cached for all time steps when $\alpha$ is
computed for~$L$.
In further experiments not shown here, we worked with a non-linear
variant of $\lambda$ reciprocally weighing-in the distance to the
nearest obstacle.

In \Cref{l:24}, we initialised our settings with a
fixed $I$ defining $N$ to be the maximum horizon
and ${\delta_I}=0$ to disable temporal scope extension.  For
$\mNFBizE[\mathcal{X}]$, we employ a fixed small symmetric padding of
2~units across all scenarios.  Overall, $\mHtiAEm{U}$ is only used
for $q\in\{\mavlbEx{depart}$, $\mavlbEx{cruise}$,
$\mavlbEx{arrive}\}$ where
$\hat\mimLwVu=\mimLwVu[fly]$.
In \Cref{l:26}, we use
$\Omega\equiv q=\mavlbEx{standby}$.  Under
$\delta$-perforation \eqref{l:12}, \textsc{updTask} is
constrained to liveness-preserving updates of $\mlYEpIk$.

Note how \Cref{l:26} leaves $\mavlbEx{standby}$, which
fulfils $\Omega$, by immediately performing the globally enabled
$\mGJEKgb{start}$ event.  In the AAV example, we keep jumps
deterministic.  Our implementation would resolve non-deterministic
jumps by taking the first available.
However, in complex applications, the discrete part of the hybrid game
will require an informed strategy as well.

To focus on the more relevant cases, \textsc{genDist} in
Line~8 only generates horizontal
disturbances.  In particular, wind (N, W, S, E) is simulated randomly,
following a semi-Markov scheme with finite memory.  In the scenarios,
wind is more likely to change gradually than spontaneously.

\begin{figure}[t]
  \centering
  \includegraphics[width=\linewidth,trim=220 880 200 0,clip]{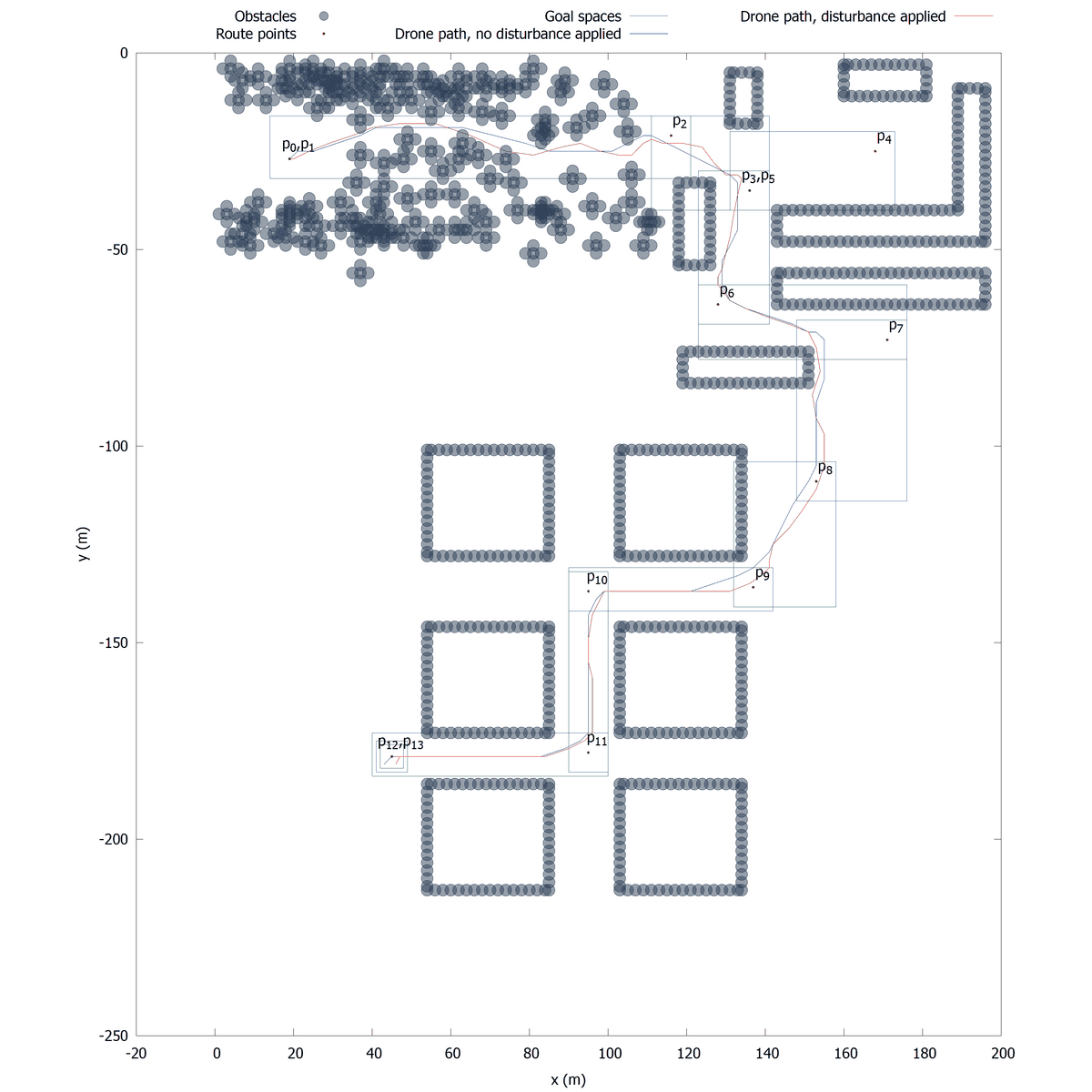} 
  \caption{Play of $G$ with ${\bar{r}}=\{{\mHtiAEm{p}}_i\}_{i=0}^n$
    for the industrial scenario ($n=13$)}
  \label{l:29}
\end{figure}

\begin{figure}
  \centering
  \includegraphics[width=\linewidth,trim=320 600 320 250,clip]{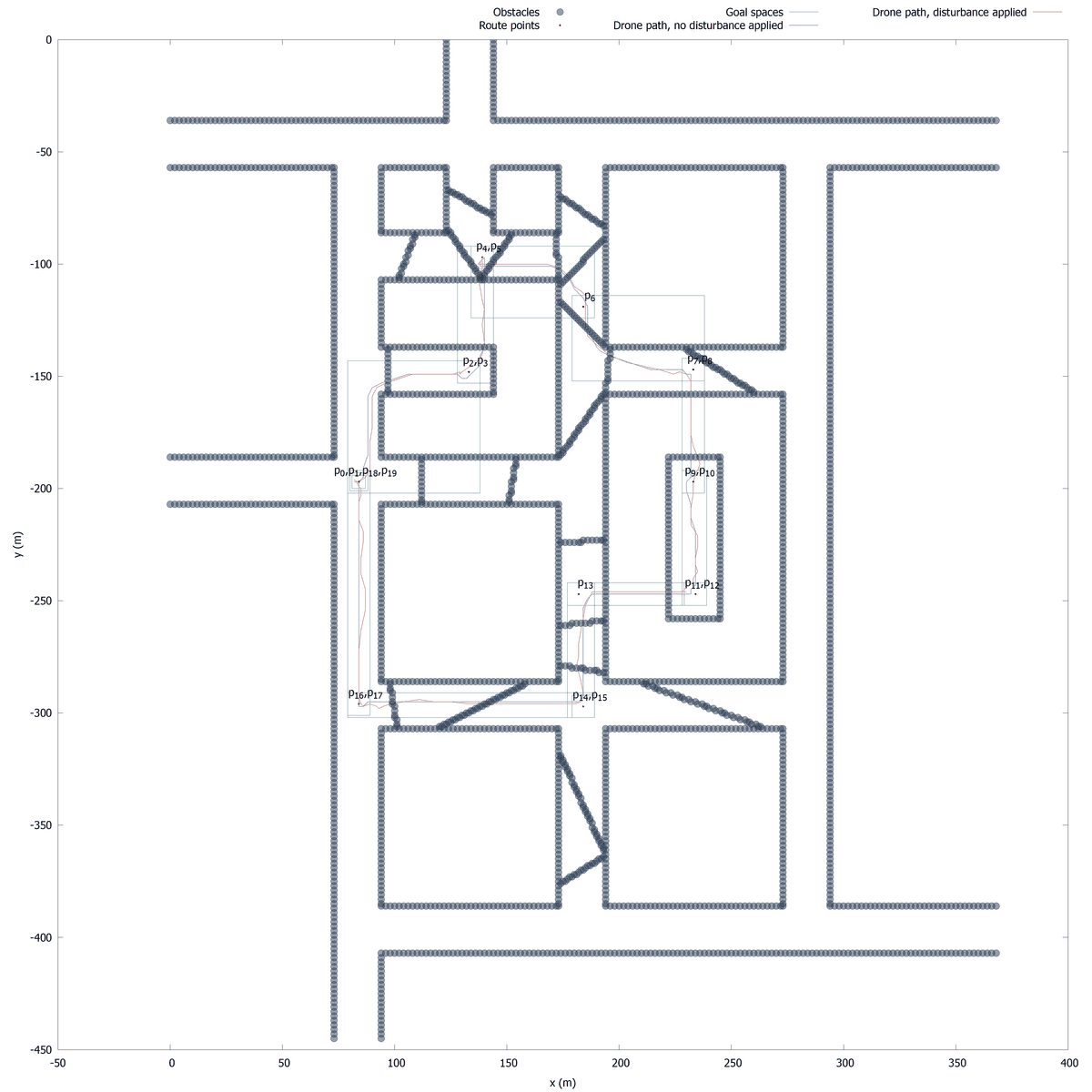} 
  \caption{Play of $G$ with ${\bar{r}}=\{{\mHtiAEm{p}}_i\}_{i=0}^n$
    for the streets scenario ($n=11$)}
  \label{l:30}
\end{figure}

\begin{figure}
  \includegraphics[width=.32\linewidth]{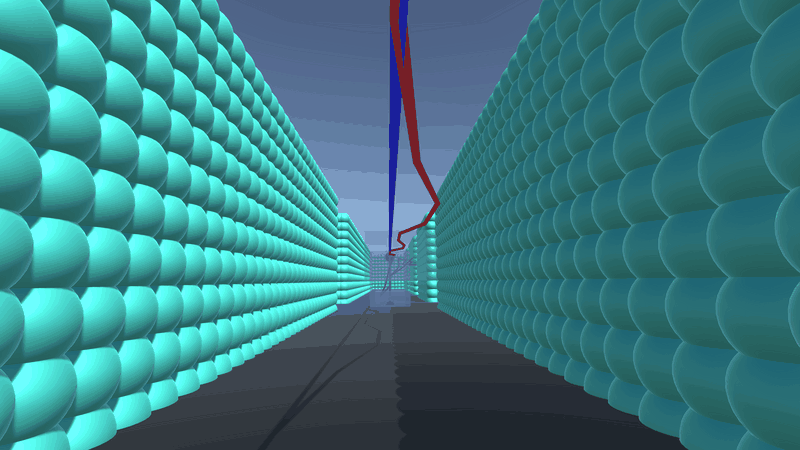}\hfill
  \includegraphics[width=.32\linewidth]{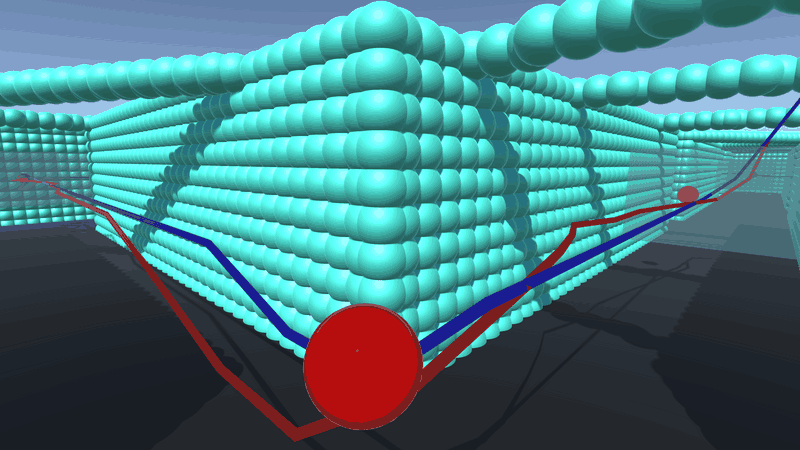}\hfill
  \includegraphics[width=.32\linewidth]{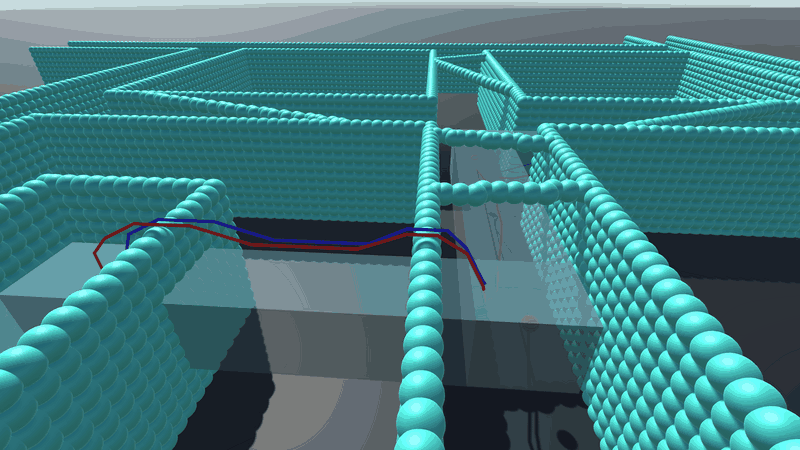}\\    
  \includegraphics[width=.32\linewidth]{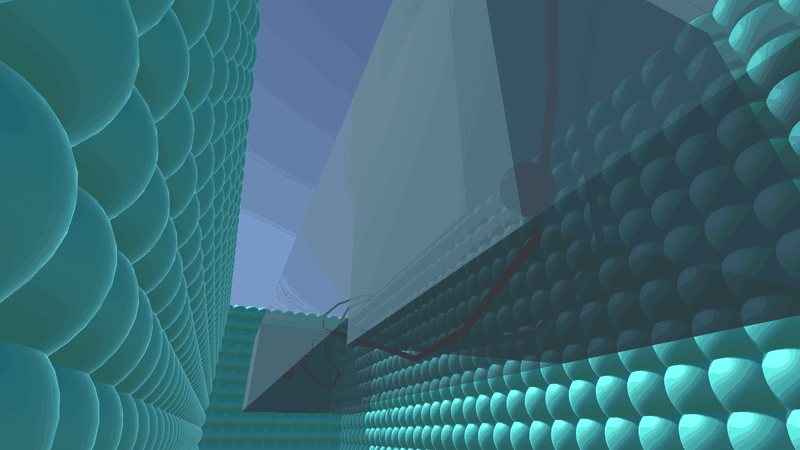}\hfill
  \includegraphics[width=.32\linewidth]{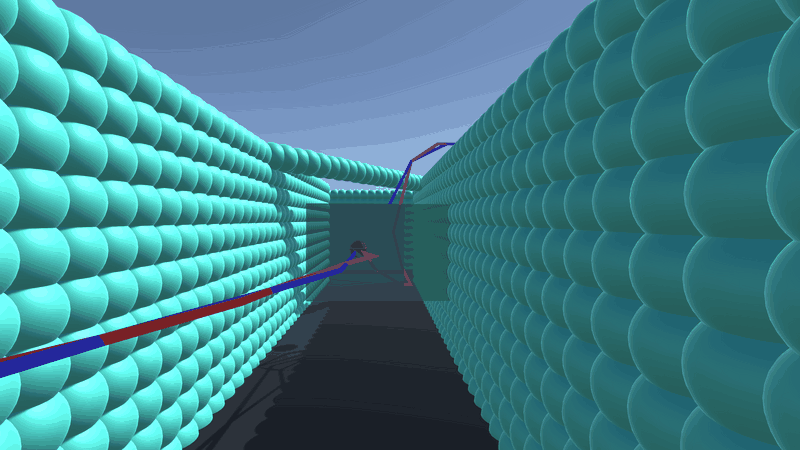}\hfill
  \includegraphics[width=.32\linewidth]{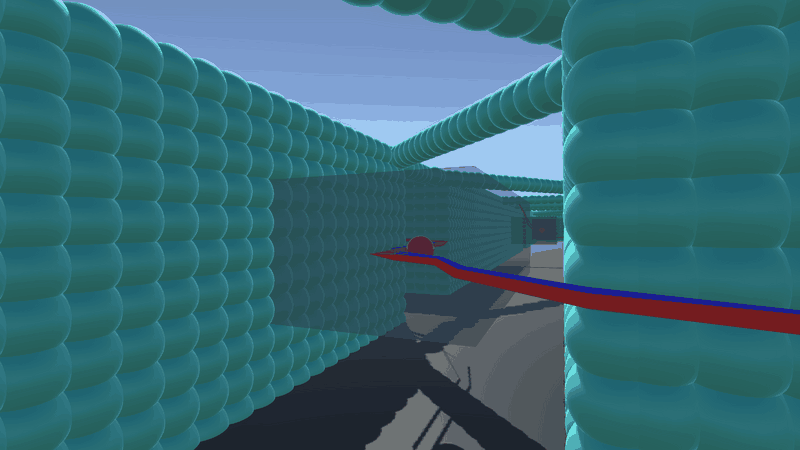}\\    
  \includegraphics[width=.32\linewidth]{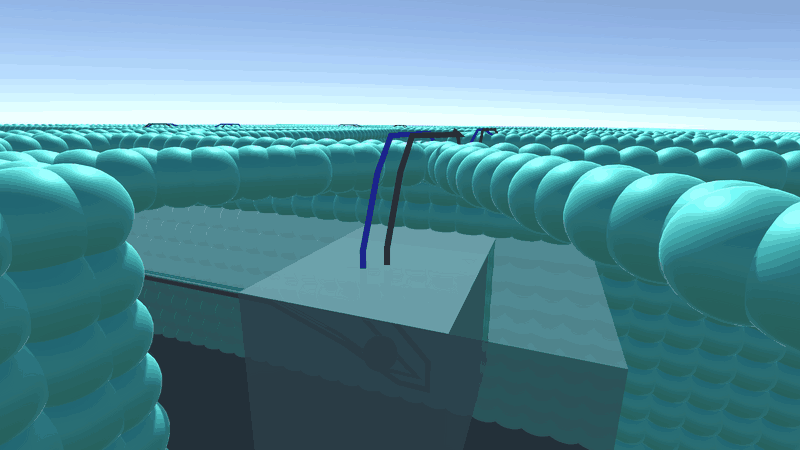}\hfill
  \includegraphics[width=.32\linewidth]{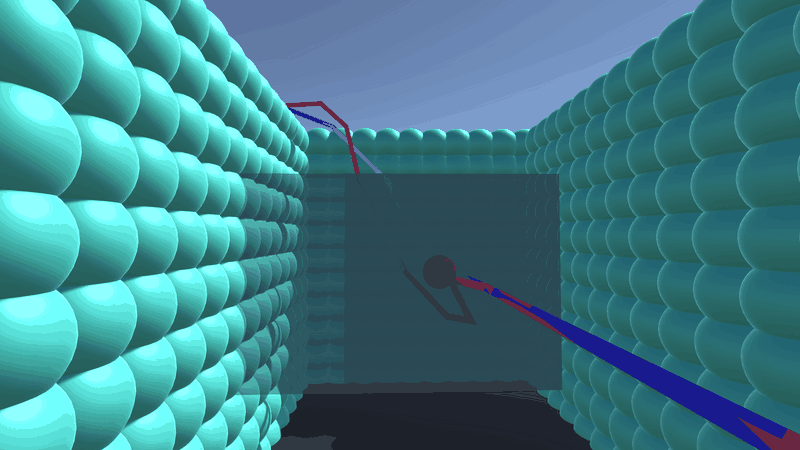}\hfill
  \includegraphics[width=.32\linewidth]{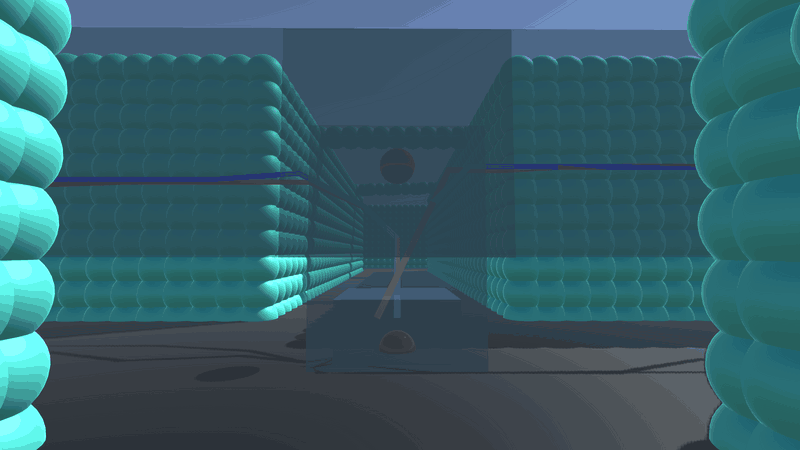}
  \caption{Snapshots of a robust play in the \emph{streets} scenario.
    In \textcolor{blue}{blue}, the near-optimal reference trajectory
    enforced by ${\mHtiAEm{u}}^*$ without disturbance being applied.  In
    \textcolor{red}{red}, a simulation of ${\mHtiAEm{u}}^*$ with random
    ${\mHtiAEm{d}}$ applied.  Transparent rectangles visualise goal
    regions.}
  \label{l:31}
\end{figure}

\paragraph{Hybrid Game Plays.}

For validating our notion of robustness, the
\Cref{l:29,l:30}
illustrate the plays of $G$ in the industrial and streets
scenarios and \Cref{l:31} provides a 3D
walkthrough of a play in the streets scenario.  A play for the yard
scenario is shown in \Cref{l:4}.
The \textcolor{blue}{blue} AAV trajectory marks the center of a
reference tube (i.e., ${\mHtiAEm{d}}=\mHtiAEm 0$) and, for comparison, the
\textcolor{red}{red} trajectory is the result of random disturbance
being applied (i.e., randomised ${\mHtiAEm{d}}$ no worse than ${\mHtiAEm{d}}^*$).
We apply the cost function from \eqref{l:28}.

The rectangle around a segment $({\mHtiAEm{p}}_{i+1},{\mHtiAEm{p}}_{i+2})$
indicates the goal region~$\rho$ of the route tracking task for segment
$({\mHtiAEm{p}}_i,{\mHtiAEm{p}}_{i+1})$.
$\rho$ circumscribes the next pair of waypoints and, thus, allows
edge cases (cf. \Cref{l:4}) where certain waypoints
(e.g.,~${\mHtiAEm{p}}_5$) are circumvented for the benefit of shortcuts (e.g.,
$({\mHtiAEm{p}}_4,{\mHtiAEm{p}}_6)$).

3D visualisation can provide valuable insights into the vertical scene
topology and the trajectory resulting from a play.  During model
validation and game design, it can help to identify parameters, such
as beneficial placement of waypoints in the context of scope shaping
or the robustness margin~$\delta$.  The 3D scenery can increase the
confidence in the robustness guarantee provided by the synthesised
controllers through a comparison of the trajectories resulting from a
worst-case play and a play with random disturbance.

\paragraph{Data from the Plays.}

\Cref{l:32} summarises key indicators of the
scenarios, such as the total run-time $t({\bar{r}})$ of the play for
route ${\bar{r}}$, time $t({\mHtiAEm{p}}_i)$ to compute ${\mHtiAEm{u}}^*$ for
segment $({\mHtiAEm{p}}_i,{\mHtiAEm{p}}_{i+1})$, and the maximum number (\#)
of states in~$\mathcal{X}$.  Moreover, the expected savings in peak
memory usage of the quasi-stationary controller are summarised in
\Cref{l:34} for comparison.

\begin{table}[t]
  \footnotesize
  \begin{tabularx}{\linewidth}{lR{.15}R{.15}R{.15}R{.15}R{.15}R{.15}R{.15}R{.15}}
    \toprule
    \emph{Scenario}
    & Play time $t({\bar{r}})$
    & Max.\ \# stages ($N$) 
    & Max.\ mem.\ usage 
    & Max.\ $\muyxJqs\mathcal{X}$
    & Avg.\ $\muyxJqs\mathcal{X}$
    & Max.\ $t$ per dDP 
    & Avg.\ $t$ per $10^6$ states 
    & Avg.\ $\frac{t({\mHtiAEm{p}}_0)}{t({\mHtiAEm{p}}_{>0})}$
    \\                 & [h:m] &    & [GiB]  & $\cdot 10^6$   & $\cdot 10^6$ & [m:s] & [s]
    \\\midrule
    Yard               & 01:39 & 30 & 20.1   & 105            &  36.4        & 25:38 & 0.597 & 1.94  \\
    Industrial         & 01:51 & 60 & 37.4   & 108            &  25.8        & 47:58 & 0.798 & 3.52  \\
    Random             & 01:03 & 30 &  5.8   &  65            &  25.7        & 06:54 & 0.626 & 1.56  \\
    Streets            & 06:43 & 50 & 36.2   & 172            &  48.7        & 51:11 & 1.110 & 4.50 
    \\\bottomrule
  \end{tabularx}
  \caption{$\delta$-robust synthesis compared by metric across the scenarios}
  \label{l:32}
\end{table}

\section{Evaluation and Discussion} 
\label{l:33}

\paragraph{Performance.}

\begin{wrapfigure}[13]{r}{.5\linewidth}
  \vspace{-2em}
  \includegraphics[width=\linewidth]{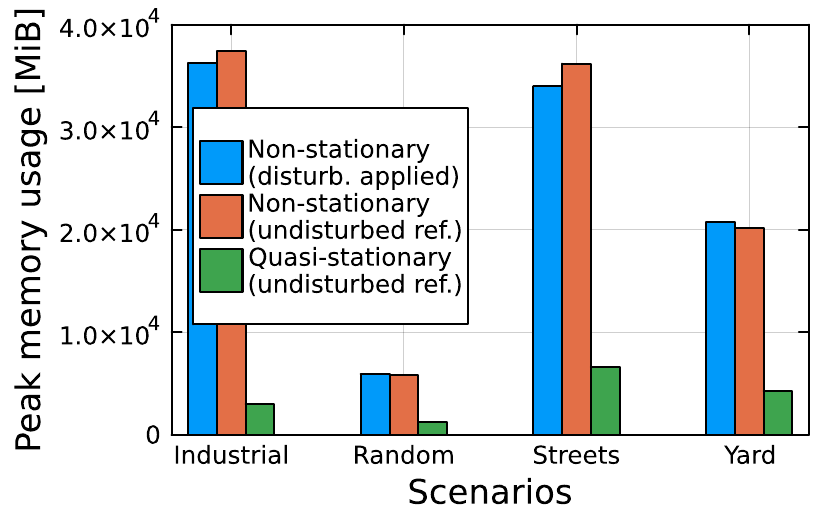}
  \caption{Peak memory usage of \Cref{l:26}
    by scenario and controller type}
  \label{l:34}
\end{wrapfigure}

Our example uses a linear approximation and a quadratic cost function
and is, thus, amenable to a solution by quadratic programming.
However, keeping our approach more widely applicable requires
significantly more time and memory than (non-linear) model-predictive control~(MPC) schemes
(\Cref{l:32}).  Clearly, the {C\texttt{++} prototype is not
ready for use in real-time settings.  Nevertheless, selective state
discretisation and omission (e.g., interpolating across state-time
dimensions, improved parameter settings) and high-performance
parallelisation can reduce average DDP run-time and space
requirements by at least two orders of magnitude, loosing optimality and precision but keeping much of the
exhaustiveness of the scenario coverage.  Such a run-time reduction
increases segment tracking speed as well as freedom in route planning.

\paragraph{Operationalisation of the Hyper-Policy.}

Lines~8 to~9
of \Cref{l:26} represent the environment part of the
control loop: Based on the observation (i.e., ${\mHtiAEm{x}}$ is an estimate
provided by an observer), control input and disturbance are applied in
parallel, followed by the next observation.  Because of this real-time
dependency,
Lines~6 to~7
need to be pre-computed to be available after the jumps.

Ideally, scenario parameters allow enough time to compute ${\mHtiAEm{u}}$
for the next segment $({\mHtiAEm{p}}_i,{\mHtiAEm{p}}_{i+1})$ based on data from local
environmental perception before reaching~${\mHtiAEm{p}}_i$.  Alternatively,
there is time to compute ${\mHtiAEm{u}}$ for $({\mHtiAEm{p}},{\mHtiAEm{p}}_i)$ before being
required to give up~${\mHtiAEm{p}}$ as a potential waiting position (e.g.,
where ${\mHtiAEm{u}}$ is a 
PID controller to robustly hold the AAV at ${\mHtiAEm{p}}$).
However, if the scenario
at hand does not permit a strong enough approximation,
$\mHtiAEm{U}$ will have to be pre-computed~(prior to or during 
flight)\footnote{Pre-computation is feasible for scope-padded segments
  $({\mHtiAEm{p}}_i,{\mHtiAEm{p}}_{i+1})$ but not arbitrary~${\mHtiAEm{s}}$.}  for a
sufficiently long prefix of~${\bar{r}}$.

\paragraph{Sound Value-Fixpoint Approximations.}

In the \Cref{l:9,l:24}, the fixpoint $\mIHrBuo[*]$,
required for ${\mHtiAEm{u}}^*$ to be stationary (i.e., applicable
independent of time, $N\approx\infty$,
\cite{b:21}) 
and optimal, is approximated.
In particular, $\widehat{\mathit{fp}}_{\mathrm{dDP}}$ replaces the ideal but
more expensive, non-simultaneous fixpoint check
$\mIHrBuo(\mathcal{X}_{\mHtiAEm{s}},k) =
\mIHrBuo(\mathcal{X}_{\mHtiAEm{s}},k+1)$
for each~$k$ in
Line~9 of \Cref{l:9}.
On the one hand, the ideal check might, due to numerical imprecision
and instability or a too small $N$, never be successful.
On the other hand, the faster check 
$\top\not\in\mIHrBuo({\mHtiAEm{x}}\oplus\mNFBizE,k)$ in
Line~7 of
\Cref{l:24}
poorly approximates $\mIHrBuo[*]$ and would turn our synthesis into
plain shortest-path finding.  The latter might, however, suffice in
some applications.

Accepting some deviation from optimality, we instead use
$\muyxJqs{\mmbQYVt(\mathcal{X}_{\mHtiAEm{s}},k)} =
\muyxJqs{\mmbQYVt(\mathcal{X}_{\mHtiAEm{s}},k+1)}$
to check in \Cref{l:9} whether the number of states stabilises,
where a robustly safe and goal-reaching trajectory for at least
$k$ steps exists.  Then, we perform
$\top\not\in\mIHrBuo({\mHtiAEm{x}}\oplus\mNFBizE,k)$ in
\Cref{l:24}.  Although
$\mmbQYVt(\mathcal{X}_{\mHtiAEm{s}},k)$ may still evolve while
preserving its cardinality, \Cref{l:26} then either
ensures ${\mHtiAEm{x}}\in \mmbQYVt(\mathcal{X}_{\mHtiAEm{s}},k)$
before leaving the solver in
Line~7 or it terminates with a failure
because of
$\mmbQYVt(\mathcal{X}_{\mHtiAEm{s}},k)=\varnothing$
in Line~10.

A comparison of the non-stationary~${\mHtiAEm{u}}^*$, obtained for
$[k,N]$ by the above termination rule, and the
quasi-stationary 
${\mHtiAEm{u}}^\infty\colon{\mathbb{X}}\to\mQghoDv$, derived via
${\mHtiAEm{u}}^\infty({\mHtiAEm{x}})={\mHtiAEm{u}}^*({\mHtiAEm{x}},k)$, indicates
(expected) improvements of ${\mHtiAEm{u}}^\infty$
(\textcolor{DarkGreen}{green}) over ${\mHtiAEm{u}}^*$
(\textcolor{blue}{blue}) in the undisturbed case
(\Cref{l:35}).  Moreover, as the
environment follows the controller (i.e., ${\mHtiAEm{u}}^*$ cannot change its
current choice based on ${\mHtiAEm{d}}$'s current choice), applying
disturbance (\textcolor{red}{red}) will have the same effect on
${\mHtiAEm{u}}^\infty$.  \Cref{l:35} (top right)
illustrates how ${\mHtiAEm{u}}^\infty$ deviates from ${\mHtiAEm{u}}^*$ to get closer
to optimality according to~\eqref{l:22}.

\begin{figure}[t]
  \framebox{
    \includegraphics[height=3.2cm,trim=0 0 0
    50,clip]{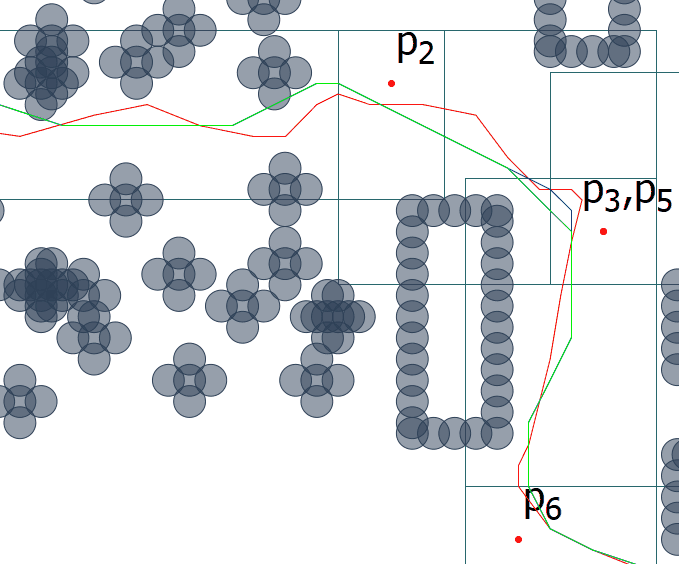}}
  \hfill
    \includegraphics[height=3.2cm,trim=0 0 0 50,clip]{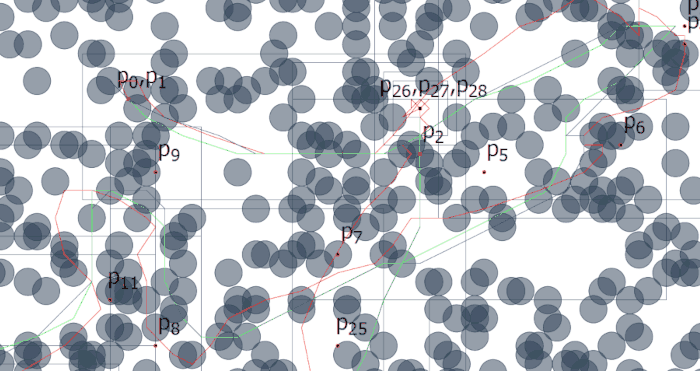}
    \\
    \includegraphics[height=3.1cm,trim=0 0 190 0,clip]{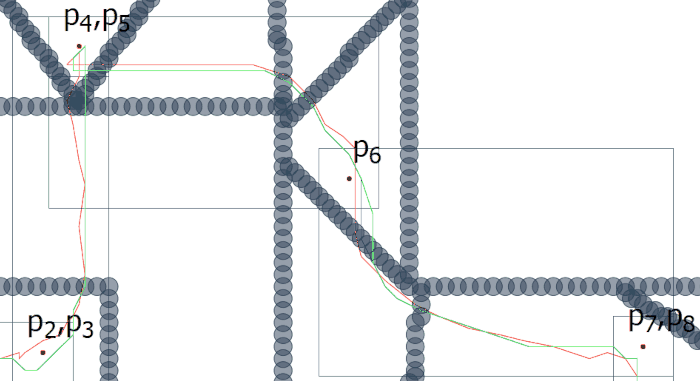}
  \hfill
  \framebox{
    \includegraphics[height=3.1cm,trim=100 0 10 0,clip]{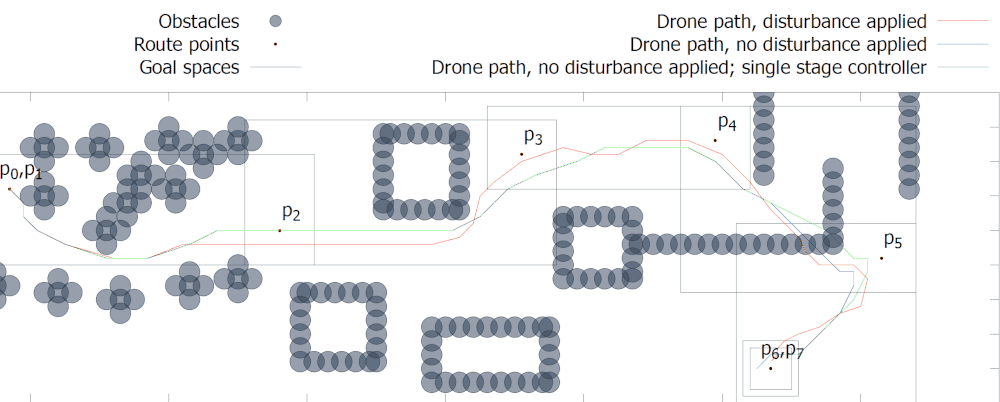}}
  \caption{Partial trajectory triples (\textcolor{blue}{non-stationary
      reference}, \textcolor{red}{random disturbance applied},
    \textcolor{DarkGreen}{quasi-stationary reference}) from the
    industrial (top left), random (top right), streets (bottom left),
    and yard (bottom right) scenarios.  The random scenario highlights
    differences between non- and quasi-stationary references.}
  \label{l:35}
\end{figure}

\paragraph{Generalisation.}

Horizon $N$, stage $k$, and alphabets $\mQghoDv$ and
$\mbPEhHP$ are largely left implicit in $\mcAFThG$, $\mIHrBuo$,
$\mmbQYVt$, $\hat\mimLwVu$, and
${\mHtiAEm{x}}^{ud}$.  Also, we employ time
resolution $1$.  However, making these parameters explicit
leads to non-essential extensions and our approach can be
parameterised by $N$ and enhanced to time-varying dynamics
and alphabets $\mQghoDv[k]$ and $\mbPEhHP[k]$ as
well as non-unit time resolutions.
Moreover, it can be extended from linear approximation (\Cref{l:9}
Line~6) to non-linear approximation of the
dynamics and to using non-linear cost-functions.

\section{Conclusion}
\label{l:36}

In this work, we specialise a conventional controller synthesis
algorithm for a parametric discretised variant of a weighted hybrid
game~$G$.  We provide a model for the integrated assurance of
robust safety, liveness, and near-optimality of controllers for
$G$ that are synthesised online, that is, during
$G$'s execution.  The abstraction we chose combines hybrid
games (e.g., reach-avoid reasoning), performance optimisation, and
numerical aspects (e.g., data sampling, quantisation) of digital
control in a way amenable to formal reasoning.  This combination
enables us to reason about robust safety and liveness guarantees of
controllers and how these guarantees impact the assurance of an
overall system.  Proofs of key statements used in this work (e.g.,
\Cref{l:26} ensures that
${\mHtiAEm{x}}\in\mmbQYVt(\mathcal{X}_{\mHtiAEm{s}},k)$; necessary
and sufficient conditions for solvability of the weighted hybrid game)
are discussed in a companion working
paper~\cite{b:6}.

Our application focuses on the modelling of
AAVs, the online synthesis of tactical controllers, and
integrating the latter with strategic and supervisory control.
Our approach can be an alternative if a combination of RL with a
shielding scheme is not desirable (e.g., lack of generalisability,
controllability, or explainability of RL's value function
approximation).

In future work, we will extend our approach to synthesise more complex
strategies and refine the integration with our
work~\cite{b:7} at the
supervisory control level.
We will improve the performance of our algorithms as suggested in
\Cref{l:33} and extend our model to be able to deal
with multiple AAVs that can form an intelligent aerial transport collective.

\bibliographystyle{splncs04}
\bibliography{main}

\begin{thebibliography}{10}
\providecommand{\url}[1]{\texttt{#1}}
\providecommand{\urlprefix}{URL }
\providecommand{\doi}[1]{https://doi.org/#1}

\bibitem{b:1}
Althoff, M., Dolan, J.M.: Online verification of automated road vehicles using
  reachability analysis. IEEE Trans. Rob.  \textbf{30}(4),  903--918 (2014).
  \doi{10.1109/TRO.2014.2312453}

\bibitem{b:2}
Bertram, J., Wei, P., Zambreno, J.: A fast markov decision process-based
  algorithm for collision avoidance in urban air mobility. IEEE Trans. Intell.
  Transp. Syst.  \textbf{23}(9),  15420--15433 (2022).
  \doi{10.1109/tits.2022.3140724}

\bibitem{b:3}
Bertsekas, D.P.: Dynamic Programming and Optimal Control, vol.~1. Athena
  Scientific, 4th edn. (2017)

\bibitem{b:4}
Clement, E., Perrin-Gilbert, N., Schlehuber-Caissier, P.: Layered Controller
  Synthesis for Dynamic Multi-agent Systems, pp. 50--68. Springer, Cham, CH
  (2023). \doi{10.1007/978-3-031-42626-1_4}

\bibitem{b:5}
Gleirscher, M.: Supervision of intelligent systems: An overview. In: Applicable
  Formal Methods for Safe Industrial Products -- Essays Dedicated to {Jan
  Peleska} on the Occasion of His 65th Birthday, LNCS, vol. 14165, pp. 1--21.
  Springer, Cham, CH (2023). \doi{10.1007/978-3-031-40132-9_13}

\bibitem{b:6}
Gleirscher, M.: Solvability of approximate reach-avoid games. CoRR  (2025).
  \doi{10.48550/arXiv.2502.04544}

\bibitem{b:7}
Gleirscher, M., Calinescu, R., Douthwaite, J., Lesage, B., Paterson, C.,
  Aitken, J., Alexander, R., Law, J.: Verified synthesis of optimal safety
  controllers for human-robot collaboration. Sci. Comput. Program.
  \textbf{218},  102809 (2022). \doi{10.1016/j.scico.2022.102809}

\bibitem{b:8}
Gu, R., Jensen, P.G., Poulsen, D.B., Seceleanu, C., Enoiu, E., Lundqvist, K.:
  Verifiable strategy synthesis for multiple autonomous agents: a scalable
  approach. Int. J. Softw. Tools Technol. Trans.  (2022).
  \doi{10.1007/s10009-022-00657-z}

\bibitem{b:9}
Henzinger, T.A.: The theory of hybrid automata. In: Verification of Digital and
  Hybrid Systems, NATO ASI Series F: Computer and Systems Sciences, vol.~170,
  pp. 265--92. Springer (2000). \doi{10.1007/978-3-642-59615-5_13}

\bibitem{b:10}
Heuillet, A., Couthouis, F., Díaz-Rodríguez, N.: Explainability in deep
  reinforcement learning. Knowledge-Based Syst.  \textbf{214},  106685 (2021).
  \doi{10.1016/j.knosys.2020.106685}

\bibitem{b:11}
H{\"o}nnecke, P.: Constrained Hybrid Optimal Control of Aerial Transport
  Systems. Master thesis, under sup. of M. Gleirscher, U{\,}Bremen (2024)

\bibitem{b:12}
Ivanov, R., Jothimurugan, K., Hsu, S., Vaidya, S., Alur, R., Bastani, O.:
  Compositional learning and verification of neural network controllers. ACM
  Trans. Embed. Comput. Syst.  \textbf{20}(5s),  1--26 (2021).
  \doi{10.1145/3477023}

\bibitem{b:13}
Kanashima, K., Ushio, T.: Finite-horizon shield for path planning ensuring
  safety/co-safety specifications and security policies. IEEE Access
  \textbf{11},  11766--11780 (2023). \doi{10.1109/access.2023.3241946}

\bibitem{b:14}
Lewis, F.L., Vrabie, D., Vamvoudakis, K.G.: Reinforcement learning and feedback
  control. IEEE Control Syst.  \textbf{32}(6),  76--105 (2012).
  \doi{10.1109/mcs.2012.2214134}

\bibitem{b:15}
Li, N., Goubault, E., Pautet, L., Putot, S.: A real-time {NMPC} controller for
  autonomous vehicle racing. In: ICACR. pp. 148--155. IEEE (2022).
  \doi{10.1109/icacr55854.2022.9935523}

\bibitem{b:16}
Mitsch, S., Ghorbal, K., Vogelbacher, D., Platzer, A.: Formal verification of
  obstacle avoidance and navigation of ground robots. Int. J. Rob. Res.
  \textbf{36}(12),  1312--1340 (2017). \doi{10.1177/0278364917733549}

\bibitem{b:17}
Schnittka, T., Gleirscher, M.: Synthesising robust strategies for robot
  collectives with recurrent tasks: A case study. In: Luckcuck, M., Xu, M.
  (eds.) FM Auton. Sys. (FMAS), 6th Workshop. EPTCS, vol.~411, pp. 109--125.
  OPA (2024). \doi{10.4204/EPTCS.411.7}

\bibitem{b:18}
Shalev-Shwartz, S., Shammah, S., Shashua, A.: Safe, multi-agent, reinforcement
  learning for autonomous driving. Tech. rep., Mobileye (2016)

\bibitem{b:19}
Taye, A.G., Valenti, R., Rajhans, A., Mavrommati, A., Mosterman, P.J., Wei, P.:
  Safe and scalable real-time trajectory planning framework for urban air
  mobility. J. Aero. Inf. Sys. pp. 1--10 (2024). \doi{10.2514/1.i011381}

\bibitem{b:20}
Thumm, J., Althoff, M.: Provably safe deep reinforcement learning for robotic
  manipulation in human environments. In: {ICRA}. {IEEE} (2022).
  \doi{10.1109/icra46639.2022.9811698}

\bibitem{b:21}
Tomlin, C.J., Lygeros, J., Sastry, S.S.: A game theoretic approach to
  controller design for hybrid systems. Proc. IEEE  \textbf{88}(7),  949--970
  (2000). \doi{10.1109/5.871303}

\end{thebibliography}
\end{document}